\documentclass[preprint,aps,pre,superscriptaddress]{revtex4-1}
\usepackage{amsmath}
\usepackage{amssymb}
\usepackage{graphicx}
\usepackage{url,hyperref}
\usepackage[usenames,dvipsnames]{xcolor}
\hypersetup{colorlinks=true, linkcolor=BrickRed, urlcolor=blue!50!black, citecolor=blue!50!black}
\usepackage[capitalize]{cleveref}
\usepackage{cleveref}
\crefrangelabelformat{equation}{(#3#1#4)$-$(#5#2#6)}
\begin{document}
\title{Influence of Roughening Transition on Magnetic Ordering}
\author{Nalina Vadakkayil}
\affiliation{Theoretical Sciences Unit and School of Advanced Materials, Jawaharlal 
Nehru Centre for Advanced Scientific Research, Jakkur P.O., Bangalore 560064, India}
\author{Sanat K. Singha}
\affiliation{Theoretical Sciences Unit and School of Advanced Materials, Jawaharlal 
Nehru Centre for Advanced Scientific Research, Jakkur P.O., Bangalore 560064, India}
\affiliation{Assam Energy Institute, Centre of Rajiv Gandhi Institute of Petroleum Technology, 
Sivasagar 785697, India}
\author{Subir K. Das}
\email{das@jncasr.ac.in}
\affiliation{Theoretical Sciences Unit and School of Advanced Materials, Jawaharlal 
Nehru Centre for Advanced Scientific Research, Jakkur P.O., Bangalore 560064, India}
\date{\today}

\begin{abstract}
In the literature of magnetic phase transitions, in addition to a critical point, existence 
of another special point has been discussed. This is related to the 
broadening of interface between two different ordering phases and is referred 
to as the point of roughening transition. 
While there exists good understanding on equilibrium properties associated with this transition, 
influence of this on nonequilibrium dynamics has not been investigated. 
In this paper we present comprehensive results, from Monte Carlo simulations, on coarsening 
dynamics in a system, over a wide range of temperature, in space dimension $d=3$, for which 
there exists roughening transition at a nonzero temperature. 
State-of-the-art analysis of the simulation data, on structure, growth and aging, 
shows that the onset of unexpected {\it{glass-like}} slow dynamics in this system, 
that has received much attention in recent times, for quenches to zero temperature, actually occurs at this 
transition point. This demonstrates an important structure-dynamics connection 
in phase-ordering dynamics. We compare the key results with those from $d=2$, 
for which there exists no non-zero roughening transition temperature. 
Absence of the above mentioned anomalous features 
in the latter dimension places our conjecture on the role of roughening transition 
on a firmer footing.
\end{abstract}
%\pacs{68.05.Cf, 64.60.Ht, 64.70.Ja, 64.70.qj, 66.10.cg}
\maketitle 
\section{Introduction}
Over past several decades there has been significant 
interest \cite{spirin1,spirin2,oliveira,kondrat,arenzon,olejarz1,olejarz2,olejarz3,corberi,blanchard,blanchard2,amar,shore,cueille,das1,das2,das3,denholm,godriche,mullick,yu,cugliandolo,liu,ohta,yeung2} 
in the understanding of ordering dynamics following quenches of paramagnetic configurations 
to the ferromagnetic region by crossing the critical temperature \cite{mfisher1} $T_c$. 
In recent times {\it{glass-like}} dynamics in a popular ordering system, 
despite the absence of in-built frustration,  
following quenches to the final temperature $T_f = 0$, drew attention \cite{olejarz1,olejarz2,olejarz3,blanchard,blanchard2,shore,cueille,das1,das2,das3}. 
This slow dynamics perhaps is due to non-conventional structure formation. 
The observation is striking and it is unknown whether such unexpected behavior \cite{corberi,das1,das2,das3} 
is specific to $T_f=0$. It is possible that the origin is at the roughening 
transition \cite{beijern}, that occurs at a much higher temperature $T_R$ $(<T_c)$, 
given that below $T_R$ interfaces are sharp. Knowledge of this is crucial not only in the understanding of 
this intriguing fact but also in establishing structure-dynamics coupling in the general context of growth phenomena. 
Interestingly, the consequences of the presence or absence of rough interfaces, of importance 
in real systems, in such nonequilibrium phenomena has not been investigated, though the 
equilibrium aspects of roughening transition is reasonably well explored. 

Some of the key aspects of ordering dynamics 
\cite{bray,puri,dfisher,yeung1,henkel,lorenz,liu,ohta,das4,yeung2,huse,das5,allen,landau} are: 
i) self-similarity and scaling property of structure, 
ii) growth of the latter, and iii) related aging. The structure is typically probed via the 
two-point equal time $(t)$ correlation function \cite{bray}, which, for a spin system, 
reads $C(r,t)= \langle S_i(t)S_j(t) \rangle-\langle S_i(t)\rangle \langle S_j(t)\rangle$, $S_i$ 
and $S_j$ representing orientations of spins or atomic magnets at sites $i$ and $j$, located $r$ 
distance apart. It is also customary to study the Fourier transform of $C(r,t)$, 
the structure factor, $S(k,t)$, $k$ being the wave number \cite{bray}. 
The latter has direct experimental relevance. 
These quantities obey certain scaling properties when the growth is self-similar. E.g., in 
simple situations, when structures are non-fractal, $C(r,t)$ satisfies \cite{bray} 
$C(r,t)\equiv\tilde C(r/\ell)$, $\ell$ being the average domain size or characteristic length scale 
of the growing system at time $t$ and $\tilde C(x)$ a time-independent master function. 
In such a situation $\ell$ is expected \cite{bray} to grow as $\sim t^\alpha$. A power-law behavior 
is expected for aging phenomena also. In the latter case 
the autocorrelation function \cite{puri,dfisher}, 
$C_\textrm{ag}(t,t_w)= \langle S_i(t)S_i(t_w)\rangle-\langle S_i(t)\rangle\langle S_i(t_w)\rangle$, 
should scale as $\sim$ $(\ell/\ell_w)^{-\lambda}$, in the asymptotic limit when $\ell >> \ell_w$. 
Here, $t_w$ $(\leq t)$ is the waiting time or age of the system and $\ell_w$ is the 
value of $\ell$ at $t=t_w$.

For uniaxial ferromagnets one expects \cite{bray,allen} $\alpha=1/2$. The structure in this case 
is supposed to be described by the Ohta-Jasnow-Kawasaki (OJK) function \cite{bray,ohta}: 
$C(r,t)=\frac{2}{\pi}\sin^{-1}[\exp(-r^2/Dt)]$, $D$ being a constant. Note that while the form 
of $C(r,t)$ and the value of growth exponent $\alpha$ are independent of space 
dimension $d$, aging exponent $\lambda$ changes with it. 
The values of $\lambda$ for this ordering, as obtained by Liu and Mazenko (LM) \cite{liu}, 
are expected to be $\simeq1.67 = \lambda_3^{\rm{LM}}$ in $d=3$, 
whereas in $d=2$, it is $\lambda_2^{\rm{LM}} \simeq 1.29$. Unless otherwise mentioned, 
in the rest of the paper all our discussions are for $d=3$. 

While these theoretical predictions were confirmed via the Monte Carlo \cite{landau} simulations 
of the Ising model, for moderately high values of  $T_f$, 
striking deviations were reported for $T_f=0$, in $d=3$. 
In the latter case, several works concluded that 
$\alpha=1/3$ or the growth is even slower. Most recently it was reported that 
the OJK function does not \cite{das3} describe the pattern at $T_f = 0$. Furthermore, 
$\lambda$ was also estimated \cite{das3} to be much weaker than $\lambda_3^{\rm{LM}}$. 
Thorough investigations, we believe, are necessary, in order to arrive at a complete and correct picture. 
It needs to be understood if such anomalies bear any connection with any other special point. 
If such a special point turns out to be that of the roughening transition, important relation 
concerning structure and dynamics \cite{das6} can be established in the nonequilibrium context. 
Our study clearly suggests that the above 
mentioned anomalous features are not specific to $T_f=0$. The onset of 
the anomalies occurs at the roughening transition. 
The key results have been verified by sophisticated finite-size 
scaling analysis \cite{das4,mfisher2,das_pre_fss}. 
We believe that these will inspire 
novel investigations, thereby explaining intriguing dynamical phenomena in the 
nonequilibrium domain.  
%------------------------------------------------------------------------------------------------
%------------------------------------------------------------------------------------------------
\section{Model and Methods}
We choose $J>0$ in the Ising Hamiltonian \cite{landau,mfisher1} $H=-J\sum_{<ij>}S_iS_j$, 
where $S_i$ and $S_j$ can take values $+1$ and $-1$, corresponding to up and down 
orientations of the atomic magnets. We study this model on a simple cubic lattice, having 
$T_c\simeq4.51J/k_B$ \cite{landau}, where $k_B$ is the Boltzmann constant. For the limited set of 
results in $d=2$, we considered the square lattice. 
Note that in this case $T_c \simeq 2.27 J/k_B$ \cite{landau}. 
In $d=3$, the value of $T_R$ for this model is $\simeq 2.57 J/k_B$ \cite{beijern}, 
whereas a non-zero roughening transition temperature does not exist in $d=2$. 

Moves in our MC simulations were tried by randomly choosing a spin and changing 
its sign \cite{landau, glauber}.  
These were accepted by following standard Metropolis criterion \cite{landau}. Unit of time 
in our simulations is a MC step (MCS) that consists of $L^d$ trial moves, $L$ being the 
linear dimension of a cubic or a square box, in units of the lattice constant. 

All our results are presented after averaging over runs with 50 independent random initial 
configurations, with $L=512$. Periodic boundary conditions were applied in all possible 
directions. Average domain lengths were measured, from the simulation snapshots, as the 
first moments \cite{das5} of the domain-size distribution function, in which length of a 
domain was estimated as the distance between two successive interfaces along any Cartesian 
direction. Results on the structure and growth were obtained after appropriately eliminating 
the thermal noise in the snapshots via a majority spin rule \cite{das5}. 
We repeat, unless otherwise stated, the results are from $d=3$.
%----------------------------------------------------------------------------------------------
%----------------------------------------------------------------------------------------------
\section{Results}
\begin{figure}
\centering
\includegraphics*[width=0.5\textwidth]{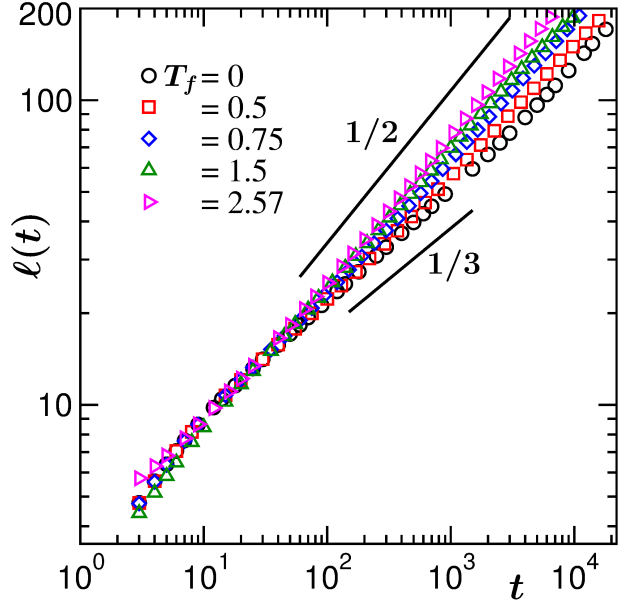}
\caption{\label{fig1}Average domain lengths, $\ell(t)$, are plotted versus 
time, on a log-log scale. Results from different final temperatures 
are presented. The solid lines are power-laws. The values of the exponents 
are mentioned.}
\end{figure}

In Fig. \ref{fig1} we show $\ell$ versus $t$ plots for several values of $T_f$. 
For $T_f=0$ few early works were suggestive of \cite{olejarz2,amar,shore,cueille} an exponent 
$\alpha=1/3$ or even slower growth. Similar quantitative behavior is seen here as well. 
However, at very late time a crossover \cite{das1,das3} of 
the exponent to a higher value, viz., $\alpha=1/2$, can be appreciated. The very early works 
could not capture this, either due to consideration of small systems or simulations over 
short periods, owing, perhaps, to inadequate computational resources. As can be seen, such 
a {\it{slow looking}} early growth is not unique to $T_f=0$. The data sets 
for nonzero $T_f$ values also exhibit similar trend. However, with the increase of $T_f$ 
departure from this slow behavior occurs earlier, the $1/3$-like regime ceasing to exist for 
$T_f = T_R = 2.57$. At this stage, it is worth warning that the early 
evolution should not be taken seriously, at the quantitative level. This is because, 
during this period satisfaction of the scaling property of the correlation function is not 
observed, as demonstrated below. 

\begin{figure}
\centering
\includegraphics*[width=0.55\textwidth]{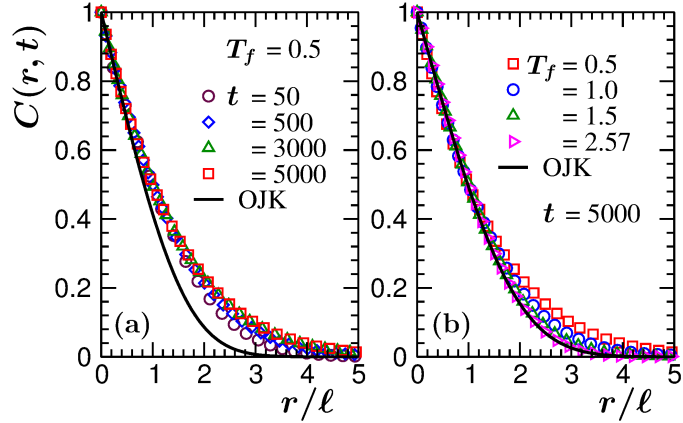}
\caption{\label{fig2}(a) Two-point equal time correlation functions, $C(r,t)$, 
are plotted versus the scaled distance $r/\ell$, for $T_f=0.5$. Data from few 
different times are shown. 
(b) Same as (a) but here we have shown $C(r,t)$ from different final temperatures. 
In each of the cases we have chosen $t=5000$ that fall in the scaling regimes. 
In both (a) and (b) the continuous lines represent the Ohta-Jasnow-Kawasaki (OJK) 
function. The correlation functions have been plotted in such a way that there 
exists good collapse in the early abscissa range.}
\end{figure}
\begin{figure}
\centering
\includegraphics*[width=0.55\textwidth]{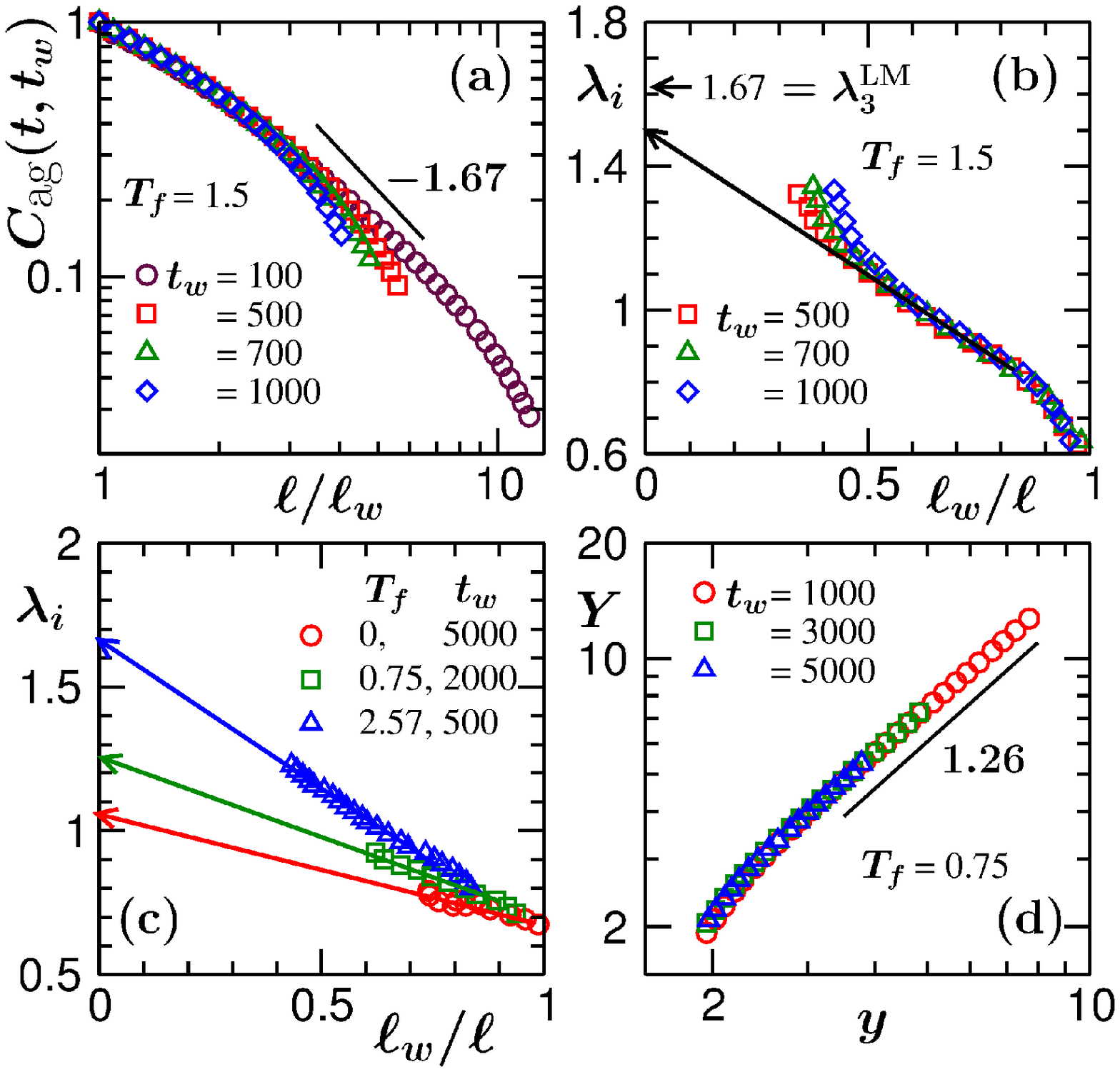}
\caption{\label{fig3}(a) $C_{\textrm{ag}} (t,t_w)$, the autocorrelation functions, 
are plotted versus $\ell/\ell_w$, on a log-log scale. Data from different $t_w$, 
for $T_f=1.5$, are included. The solid line is a power-law with 
$\lambda = \lambda_3^{\rm{LM}} = 1.67$. 
(b) The instantaneous exponents $\lambda_i$ are shown as a function of $\ell_w/\ell$, 
for the same final temperature. The arrow-headed line is a linear extrapolation 
to the $\ell/\ell_w = \infty$ limit, done by excluding the late time finite-size 
affected as well as early time domain-magnetization relaxation parts. 
(c) Here we have shown $\lambda_i$, as a function of $\ell_w/\ell$, 
for a few different values of $T_f$. In each of the cases $t_w$ belongs to the 
scaling regime. The arrow-headed lines are linear guides to the eyes. 
(d) Finite-size scaling plot of $C_{\rm{ag}}(t,t_w)$ for $T_f = 0.75$. 
The solid line there corresponds to a power-law with the value of the exponent 
mentioned near the line.}
\end{figure}
In Fig. \ref{fig2}(a) we show plots of $C(r,t)$, from different times, by scaling the 
distance axis by a characteristic length, extracted by exploiting the satisfaction of 
scaling of $C(r,t)$ at small distances, for $T_f=0.5$. 
It appears that the collapse starts only from 
$t \gtrsim 1000$, approximately the time since when departure to $\alpha=1/2$ behavior starts. 
This general picture is true for other low temperatures also. In Fig. \ref{fig2}(b) we have 
shown $C(r,t)$, again versus $r/\ell$, from the scaling regimes of different $T_f$ values. 
Interestingly, $\tilde{C}(r/\ell)$ at different $T_f$ values do not agree with each other. 
However, with the increase of $T_f$ the agreement with the OJK function 
$[\tilde{C}_{\rm{OJK}} (r/\ell)]$ \cite{ohta} keeps getting better. 
This observation suggests that perhaps there exists a special temperature $T_{\rm{sp}}$ ($<T_c$), 
beyond which the coarsening  dynamics is more unique than below it. 
In fact for $T_f = T_R$ the agreement between simulation data and the OJK function is quite well. 
We will return to this central theme after discussion of the basic results on aging. 

Fig. \ref{fig3}(a) shows plots of $C_\textrm{ag}(t,t_w)$, with the variation of $\ell/\ell_w$. 
The value of $T_f$ for this representative case is set at $1.5$. Data sets from a few 
different waiting times are shown. 
Good collapse of data is visible for the considered values of $t_w$. However, there exist 
deviations from the master curve, for $\ell/\ell_w>>1$. These are related 
to finite-size effects \cite{das4}. Decay in the finite-size unaffected regime does not appear 
consistent with the LM value \cite{liu} -- see the disagreement with the solid line. 
In Fig. \ref{fig3}(b) we show the instantaneous exponent \cite{dfisher,das4} $\lambda_i$ 
$[=-\textrm{d}\ln C_{\textrm{ag}}(t,t_w)/\textrm{d}\ln(\ell/\ell_w)]$ as a function of 
$\ell_w/\ell$, for multiple choices of $t_w$ lying in the scaling regime. The data sets 
appear linear in the finite-size unaffected regimes. Note that at early time 
relaxation of domain magnetization interferes and should be discarded from the process of 
estimation of $\lambda$. A linear extrapolation to $\ell/\ell_w=\infty$ provides 
$\lambda \simeq 1.5$. Given that the above quoted number lies between $\lambda_3^{\rm{LM}}$ 
and $\lambda(T_f = 0)$, being significantly different from each of these, one gets a 
strong hint on the presence of a special point. 

In Fig. \ref{fig3}(c) we show plots of $\lambda_i$, as a function of $\ell_w/\ell$, for 
few different values of $T_f$. Here we discarded the parts corresponding to finite-size effects 
and equilibration of domain magnetization. Furthermore, in each of the cases the results are from 
well inside the scaling regimes of $t_w$. 
The arrow-headed lines are related to the estimations of the values of $\lambda$, 
from linear extrapolations to the $\ell_w/\ell = \infty$ limit. Clearly, $\lambda$ depends 
strongly on $T_f$. The accuracy of these estimates is validated by the independent quantifications 
of $\lambda$ via a finite-size scaling method \cite{das4,das_pre_fss}. 
A representative exercise related to this is 
shown in Fig. \ref{fig3}(d), for $T_f = 0.75$. In this figure $Y$ is a $t_w$-independent 
scaling function and $y$ is  a dimensionless scaling variable. Note that here we have avoided 
studying systems of different sizes, contrary to the standard practice in the literature of 
such analysis. Instead, we have 
obtained collapse of data from different $t_w$ values. Note that when $t_w$ is varied a system 
has different effective sizes to grow further. Details of the scaling construction is 
provided below. 

\begin{figure}
\centering
\includegraphics*[width=0.55\textwidth]{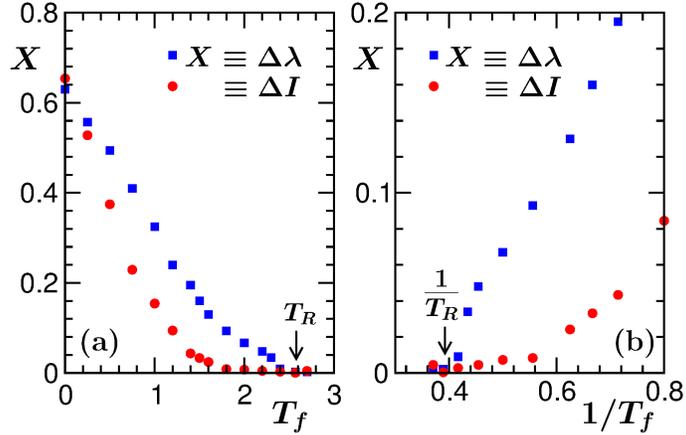}
\caption{\label{fig4}(a) Plots are shown by comparing $\Delta \lambda$ 
and $\Delta I$ with the variation of $T_f$. 
(b) Plots of $\Delta \lambda$ and $\Delta I$ as a function 
of $1/T_f$. The locations of $T_R$ are marked inside the frames.}
\end{figure}
The behavior of $\lambda_i$ in Fig. \ref{fig3}(b) and Fig. \ref{fig3}(c) 
suggest $\lambda_i = \lambda - B/x$, with $x=\ell/\ell_w$ and $B$ being a constant, 
in the finite-size unaffected late time regime. This leads to a form \cite{das4,das_pre_fss} 
$C_{\textrm{ag}} (t,t_w) = A e^{-B/x}x^{-\lambda}$. 
By taking $y = L/\ell$ as a scaling variable and $y_w = L/\ell_w$, a finite-size scaling 
function can be written as \cite{das3} $Y = C_{\textrm{ag}} (t,t_w) e^{By/y_w} y_w^\lambda$, 
where $Y$ contains a factor $y^\lambda$. When results from different $t_w$ are plotted, 
for optimum choices of the unknown parameters, including $\lambda$, there will be collapse 
of data sets that will satisfy the expected $y^\lambda$ behavior at large $y$. 
This is demonstrated in Fig. \ref{fig3}(d) for $T_f = 0.75$. 
Here the collapse is obtained for $\lambda = 1.26$, the number being 
consistent with the value that was suggested by the exercise in Fig. \ref{fig3}(c). 
Next we quantify the special temperature from a more systematic study.
\begin{figure}
\centering
\includegraphics*[width=0.55\textwidth]{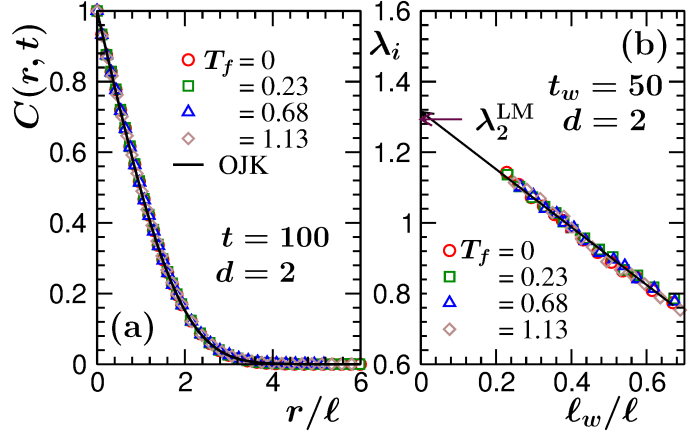}
\caption{\label{fig5}(a) Plots of scaled two-point equal time correlation 
functions, $C(r,t)$, from $d=2$, for different $T_f$ values. The time has been 
chosen from the scaling regime. The continuous line represents the OJK function.
(b) The instantaneous exponents, $\lambda_i$, are plotted as a function 
of $\ell_w/\ell$, for different $T_f$ values. The solid line is a 
common linear extrapolation of the data sets. The arrow-headed horizontal 
line points to the LM value of $\lambda$ in $d=2$.}
\end{figure}

In Fig. \ref{fig4}(a) we show $\Delta \lambda = \lambda_3^{\rm{LM}} -\lambda(T_f)$, as a 
function of $T_f$. Given that the general expectation is $\lambda_3^{\rm{LM}}$, it is 
meaningful to look at the stated difference. There appears to be a nice convergence of the 
data set to zero as $T_f \rightarrow T_{\rm{R}} \simeq 2.57$. With respect to the deviation 
of $\tilde{C}(r/\ell)$ from the OJK form \cite{ohta}, that we observed above, there may also 
be a similar trend. In this case an appropriate quantity to consider 
is $\Delta I = \int dr [\tilde{C}(r/\ell) - \tilde{C}_{\rm{OJK}} (r/\ell)]$. 
Here note that there exists an LM form for $C(r,t)$ as well \cite{liu}. However, this practically 
overlaps with the OJK function. It will be interesting to see if $\Delta I$ approaches zero 
at the same $T_f$ as in the case of $\Delta \lambda$. Thus, in Fig. \ref{fig4}(a) we have 
included the $T_f$-dependence of $\Delta I$ as well. The presented data sets are in nice 
agreement with each other, over a wide range of temperature, 
within a factor -- see Fig. \ref{fig4}(b) for a clearer convergence to zero for $T_f \rightarrow T_R$.

In Fig. \ref{fig5} we show analogous results from $d=2$. Fig. \ref{fig5}(a) contains 
results for the scaled $C(r,t)$ and in Fig. \ref{fig5}(b) we have shown data for $\lambda_i$, 
as a function of $\ell_w/\ell$. For each of the cases results from a wide range of $T_f$ are 
included. The anomalies present in $d=3$ are clearly absent in this case. 
No detectable $T_f$-dependence can be observed. The theoretical expectations are satisfied 
over the whole range of $T_f$. 
Recall that in this dimension, a nonzero roughening transition temperature does not exist 
for this model.
%-------------------------------------------------------------------------------------------
\section{Conclusions}
From extensive Monte Carlo simulations \cite{landau} we have presented results 
on nonequilibrium dynamics in the Glauber \cite{landau,glauber} Ising model. 
This mimics ordering in uniaxial ferromagnets. Our quantitative analysis of data 
from space dimension $d=3$ on structure, growth and aging,  over a wide range of 
temperature below the critical point, suggests that the low temperature behavior 
is anomalous. 

We show that that the anomalies are not unique to the case of zero temperature quench, 
as was previously thought. Various quantities exhibit zero-temperature-like trend till 
a certain nonzero value of $T_f$. Above this temperature, behavior of all the aspects 
become consistent with various theoretical expectations \cite{bray,liu,ohta,allen}. 
This {\it{transition or special temperature}} coincides with that of the roughening 
transition \cite{beijern}. Such a conclusion appears more meaningful from the fact that 
these anomalies are absent in $d=2$ and for this dimension roughening transition temperature is zero.

To understand the anomalies, more detailed theoretical investigations by exploiting the 
well travelled auxiliary field ansatz should be carried out \cite{yeung}. 
Recently, it was shown \cite{mpemba_pccp} that the ordering dynamics in the Glauber Ising 
model exhibits Mpemba effect. It needs to be further investigated in light of the present observation. 

SKD acknowledges previous collaborations on similar matter with S. Chakraborty, S. Majumder and J. Midya. 
The authors are grateful to SERB, DST, INDIA for support via Grant No. MTR/2019/001585.

\end{document}